\documentstyle{elsart}
\begin{document}

\input epsf.sty

\renewcommand{\theequation}{\thesection.\arabic{equation}}
\def\ade{$A$--$D$--$E$\space}
\def\weight#1#2#3#4#5{#1\!\!\left(\matrix{#5&#4\cr #2&#3\cr}\right)}

\def\wt#1#2#3#4#5#6{#1\!\!\mbox{
$\left(\matrix{#5&#4\cr#2&#3\cr}\biggm|\mbox{$#6$}\right)$}}

\def\wtm#1#2#3#4#5{#1\!\!\mbox{
$\left(\matrix{#5&#4\cr#2&#3\cr}\right)$}}

\def\wtd#1#2#3#4#5{#1\!\!\mbox{
$\left( #5 \,\, \matrix{ #4\cr \cr #2\cr} \,\, #3 \right)$}}

\newcommand{\hs}[1]{\hspace*{#1cm}}
\newcommand{\vs}[1]{\vspace*{#1cm}}

\newcommand{\Kp}[3]{K_{+}\biggl(\matrix{#2\vs{-0.3}&\cr&
  \hs{-0.3}#1\vs{-0.3}\cr#3}\biggr)}

\newcommand{\K}[4]{K_{#1}\biggl(\!\matrix{&#3\vs{-0.3}\cr\!\!
  #2\hs{-0.3}\vs{-0.3}&\cr&#4}\biggr)}

\def\half {\mbox{$\textstyle {1 \over 2}$}}
\def\mat {\pmatrix}
\def\smat#1{\mbox{\small $\mat{#1}$}}
\def\ba{\begin{array}}
\def\ea{\end{array}}
\def\be{\begin{eqnarray}}
\def\ee{\end{eqnarray}} 
\def\beq{\begin{equation}}
\def\eeq{\end{equation}} 
\def\no{\nonumber}
\def\lam{\lambda}
\def\ffrac#1#2{\mbox{\small $\frac{#1}{#2}$}}
\def\I{{\rm i}}

\begin{frontmatter}
\title{Surface operator content of the  $A_L$ face models}

\author{Murray T. Batchelor\thanksref{ARC}}

\address{Department of Mathematics, School of Mathematical
Sciences,\\ The Australian National University, Canberra ACT 0200,
Australia}

\thanks[ARC]{Supported by the Australian Research Council}

\begin{abstract}
A set of fixed boundary weights for both the critical dense and 
dilute $A_L$ face models is constructed from the known boundary 
weights of the related loop model. The surface operator content
and the conformal partition functions then follow from the
results obtained via the Bethe equations. 
\end{abstract}

\end{frontmatter}

\section{Introduction}
\setcounter{equation}{0}

The remarkable interplay between critical systems and the symmetries 
of conformal and modular invariance has been well illustrated
by direct calculations on exactly solved models \cite{ISZ,Cb,CH}.
One well known family of such are the critical 
$A$-$D$-$E$ models which are built from the Dynkin diagrams
of the simply laced $A$-$D$-$E$ Lie algebras \cite{Pas}. These
models can all be mapped onto an underlying loop model \cite{Pas,OBa}.
Another family of critical
models, the dilute $A$-$D$-$E$ lattice models \cite{WNS:92,Roche:92,WNS:93},
can be mapped onto the dilute O($n$) loop model \cite{N:90a,WN:93}.
The bulk operator content and the modular invariant partition functions
of these models are now well understood (see, e.g., \cite{OP}). However, 
the situation is not 
so clear for the surface operator content, at least from the perspective
of exactly solved models.

\begin{figure}[htb]
\centerline{
\epsfxsize=2.5in
\epsfbox{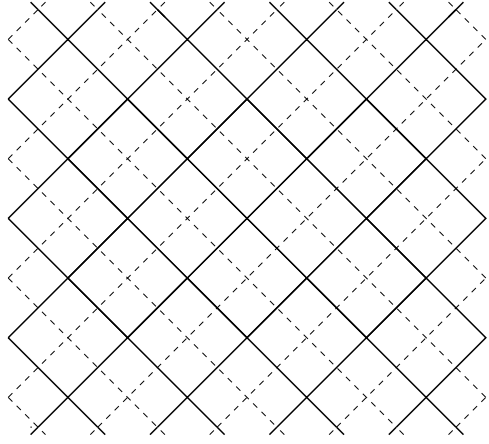}}
\caption{An open square lattice in the vertical strip geometry. 
The face models under consideration are defined on the solid lines. 
The underlying loop and vertex models are defined on the dashed lines.}
\label{lattice}
\end{figure}

The particular geometry of interest here is that shown in Fig.~\ref{lattice}.
We confine our attention to the $A_L$ models. To distinguish
between the two families, we refer to them as the dense $A_L$ and the
dilute $A_L$ models, the nomenclature arising naturally from the
corresponding loop models. 
 
\subsection{Face models}

The bulk face weights of the dense $A_L$ models at criticality 
are \cite{Pas,ABF}
\be
& & \wtd Wabcd
= \rho_8(u) \,
\delta_{a,c} A_{a,b} A_{a,d} +
       \sqrt{{S_a S_c \over S_b S_d}} \, \rho_9(u) \, \delta_{b,d} A_{a,b}
A_{b,c} \label{denseW} 
\ee
where 
\be
\rho_8 (u) =1, \qquad \rho_9 = {\sin u \over \sin (\lam - u)} \label{dense}
\ee
and $S_a = \sin a\lam$.
The variable $u$ is the spectral parameter and $\lam = \pi/h$ is the
crossing parameter with Coxeter number $h=L+1$. We are interested here
in the boundary between regimes III and IV for which $0 < u < \lam$.
The elements of the
adjacency matrix $A$ are given by
\beq
A_{a,b}=\cases{1, &$b = a \pm 1$, \cr
               0, &otherwise. }
\eeq

The bulk face weights of the dilute $A_L$ models at criticality
are \cite{WNS:92,Roche:92,WNS:93}
\be
& & \wtd Wabcd
= \;\rho_1(u)
  \delta_{a,b,c,d} +\rho_2(u) \delta_{a,b,c} A_{a,d}+\rho_3(u)
\delta_{a,c,d} A_{a,b} \no\\*
& &\quad \mbox{} +\sqrt{S_a \over S_b} \, \rho_4 (u) \delta_{b,c,d} A_{a,b}
     +\sqrt{S_c \over S_a}\,\rho_5(u) \delta_{a,b,d} A_{a,c}
     +\rho_6(u) \delta_{a,b} \delta_{c,d} A_{a,c}  \label{adeface-d}\\
& &\quad \mbox{}  +\rho_7(u) \delta_{a,d} \delta_{c,b} A_{a,b} +\rho_8(u)
\delta_{a,c} A_{a,b} A_{a,d} +
       \sqrt{{S_a S_c \over S_b S_d}}\, \rho_9(u) \delta_{b,d} A_{a,b}
A_{b,c} \no \label{diluteW} 
\ee
in which 
\be
\rho_1 (u)&=&1 + {\sin u \sin (3\lam-u)
                        \over \sin 2\lam\sin 3\lam}  \no\\
\rho_2 (u)&=&\rho_3 (u) = {\sin (3\lam-u)\over \sin 3\lam}\no \\
\rho_4 (u)&=&\rho_5 (u) = {\sin u \over \sin 3\lam} \no \\
\rho_6 (u)&=&\rho_7 (u) = {\sin u \sin (3\lam-u) \over \sin
                                     2\lam\sin 3\lam}  \\
\rho_8 (u)&=&{\sin (2\lam-u) \sin (3\lam-u)\over\sin
                                     2\lam\sin 3\lam} \no \\
\rho_9 (u)&=&-{\sin u \sin(\lam-u) \over \sin 2\lam\sin 3\lam} \no
\ee
and the generalised Kronecker delta is unity if all its arguments take the 
same value and zero otherwise. 
For these models
\begin{eqnarray}
A_{a,b}=\cases{1, &$b = a$ or $a \pm 1$, \cr
               0, &otherwise. }  
\end{eqnarray}
The regimes of interest here are 
\be
&\mbox{regime 1}
\hs{1.0} 0<u< 3\lam &\hs{0.7}\lam={\pi\over 4}{L\over L+1} \hs{0.7}
                L=2,3,\cdots\no\\
&\mbox{regime 2}
\hs{1.0} 0<u< 3\lam &\hs{0.7}\lam={\pi\over 4}{L+2\over L+1}\hs{0.7}
                  L=3,4,\cdots\no
\ee
where
\beq
h=\cases{L+2, &regime 1, \cr
         L+1, &regime 2. }
\eeq 

\begin{figure}[htb]
\centerline{
\epsfxsize=3.5in
\epsfbox{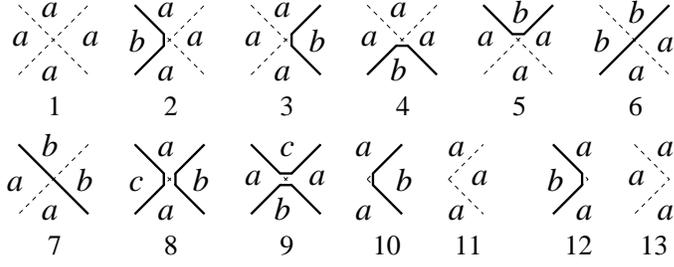}}
\caption{The allowed vertices of the loop model along with the possible
height configurations.}
\label{vertices}
\end{figure}

\subsection{Loop models}

To obtain a set of integrable boundary weights for the face models
we consider the corresponding loop models, which have
been solved in the open geometry of Fig.~\ref{lattice}~\cite{ob89,yb95a}.
The partition function of the dense O($n$) loop model on the dashed 
lattice of  Fig.~\ref{lattice} is defined by
\be
Z = \sum_{{\cal G}} \rho_8^{m_8} \rho_9^{m_9}
                    \rho_{10}^{m_{10}} \rho_{12}^{m_{12}} \; n^{P},
\ee
where the bulk weights $\rho_8$ and $\rho_9$ are given in (\ref{dense}).
The sum is over all configurations
${\cal G}$ of non-intersecting closed
loops covering all lattice bonds.
The two possible loop configurations at a vertex are
numbers 8 and 9 in Fig.~\ref{vertices}.
In configuration ${\cal G}$,
$m_i$ is the number of occurrences of the
vertex of type $i$, while $P$ is the
total number of closed loops of fugacity $n = 2 \cos\lam$.
The boundary weights are
\be
\rho_{10} = \rho_{12} = 1.
\ee

The partition function of the dilute O($n$) model is defined by 
\beq
Z = \sum_{{\cal G}} \rho_1^{m_1} \cdots \rho_{13}^{m_{13}} \; n^{P},
\label{Z}
\eeq
where now the loops need not cover all lattice bonds
and the loop fugacity is given by $n=-2\cos 4\lam$. 
The possible loop configurations at a vertex are 
shown in Fig.~\ref{vertices}, with a vertex of type
$i$ carrying a Boltzmann weight $\rho_i$. 
The bulk weights are as given in (1.5).
The integrable boundary weights are
\be
&&\rho_{10}=\rho_{12}=\sin[\ffrac{1}{2}(3\lam-u)+\epsilon],  \cr
&&\rho_{11}=\rho_{13}=\sin[\ffrac{1}{2}(3\lam+u)+\epsilon].
\label{dbw}
\ee
There is one set for $\epsilon =0$, the other for $\epsilon = \frac{\pi}{2}$.

Note that just as there is a loop formulation of the Yang-Baxter
equation, ensuring integrability of the loop model in the bulk \cite{WNB},
there is also a loop formulation of the reflection equations,
ensuring integrability at a boundary \cite{unpub}. Rather than
seeking explicit solutions of these equations our 
approach here is via the loop-to-face correspondence.

\section{Boundary face weights and exact solutions}
\setcounter{equation}{0}

The face model transfer matrix ${\bf t}_D(u)$ acts
between three diagonal rows of the lattice, labelled by the
heights $\{a_0,a_1,\ldots,a_N\}$, 
$\{b_0,b_1,\ldots,b_N\}$ and $\{c_0,c_1,\ldots,$ $c_N\}$,
with $a_{2k}=b_{2k}$ and $b_{2k-1}=c_{2k-1}$.
Here $N$ is the number of bulk edges in a row, which we
take to be even. Thus
\be
{\bf t}_D(u) = &&\Kp {b_1}{c_0}{b_0} 
\prod_{k=1}^{N/2-1}
\wtd W{a_{2k-1}}{a_{2k}}{b_{2k-1}}{a_{2k-2}} 
\wtd W{b_{2k}}{b_{2k+1}}{c_{2k}}{b_{2k-1}} 
\cr
&& \times \,\, \wtd W{a_{N-1}}{a_{N}}{b_{N-1}}{a_{N-2}}
\K{-}{\,\, b_{N-1}}{c_N}{b_N} \, . 
\ee
The bulk face weights of the $A_L$ models are as given in 
(\ref{denseW}) and (1.4).
Now consider the boundary face weights. 

\subsection{Dense $A_L$ model} 

In accord with Fig.~\ref{vertices} we define the boundary face weights
of the dense $A_L$ model to be 
\beq
\Kp {a\pm1}aa = \rho_{10}\, , \qquad 
\K{-}{\,\, a\pm1}aa = \rho_{12}\, ,
\eeq
where $\rho_{10}$ and $\rho_{12}$ may be arbitrary height-independent 
functions. General boundary face weights for this model have been
discussed in \cite{BPO,AK}. 

The diagonal-to-diagonal transfer matrix ${\bf t}_D(u)$ for the underlying
loop model has eigenvalues \cite{ob89,yb95a}
\begin{eqnarray}
\Lambda(u) = \rho_{10} \, \rho_{12} 
  \prod_{j=1}^m \frac{\sinh[u_j+ \half \I (u+\lambda)]
   \sinh[u_j- \half \I (u+\lambda)]}
  {\sinh[u_j+ \half \I (u-\lambda)]
   \sinh[u_j- \half \I (u-\lambda)]}\; ,
\label{eigs} 
\end{eqnarray}
where $u_j$ $(j=1,2,\ldots,m)$ are roots of the Bethe equations
\be
\left[\frac{\sinh[u_j+ \half \I (u-\lambda)]
    \sinh[u_j- \half \I (u+\lambda)]}
   {\sinh[u_j+ \half \I (u+\lambda)]
    \sinh[u_j- \half \I (u-\lambda)]}\right]^N \cr 
= \prod_{\stackrel{k=1}{~\neq j}}^m \frac{\sinh(u_k + u_j - \I \lambda)
    \sinh(u_k - u_j - \I \lambda)}
    {\sinh(u_k + u_j + \I \lambda)\sinh(u_k - u_j + \I \lambda)}\;
.\label{denseBE} 
\ee

\subsection{Dilute $A_L$ model} 

Again in accord with Fig.~\ref{vertices} we define the boundary 
face weights of the dilute $A_L$ model to be 
\be
\Kp {a\pm1}aa = \rho_{10}\, , \qquad \Kp aaa &=& \rho_{11} \, , \cr
\K{-}{\,\,a\pm1}aa = \rho_{12}\, , \qquad \K{-}aaa &=& \rho_{13} \, ,
\ee
where the functions $\rho_{10}$--$\rho_{13}$ are as given in
(\ref{dbw}). The origin of the $\epsilon$ factor is seen to be 
the double periodicity of the more general elliptic 
boundary weights \cite{BFZ}.

With the given normalisation of the weights, the
eigenvalues are \cite{yb95a}  
\begin{equation}
\Lambda_D(u) = \rho_8^{N-1} \rho_{10} \, \rho_{12} 
\prod_{j=1}^m \frac{\sinh(u_j + \I \lambda + \half \I u)
                    \sinh(u_j - \I \lambda - \half \I u)}
            {\sinh(u_j - \I \lambda + \half \I u)
             \sinh(u_j + \I \lambda - \half \I u) }
\end{equation}
where the $m$ roots $u_j$ satisfy 
\begin{eqnarray}
\lefteqn{
\left[\frac{\sinh(u_j - \half \I \lambda + \I \epsilon)}
           {\sinh(u_j + \half \I \lambda + \I \epsilon)}\right]^2
 \left[\frac{\sinh(u_j - \I \lambda - \half \I u)
             \sinh(u_j - \I \lambda + \half \I u)}
            {\sinh(u_j + \I \lambda - \half \I u)
             \sinh(u_j + \I \lambda + \half \I u)}
\right]^N} \hspace{90pt} \nonumber\\
&= & \prod_{\stackrel{k=1}{~\neq j}}^m 
  \frac{\sinh(u_j+u_k-2 \I \lambda)
        \sinh(u_j-u_k-2 \I \lambda)}
       {\sinh(u_j+u_k+2 \I \lambda)
        \sinh(u_j-u_k+2 \I \lambda)} \nonumber\\
& & \quad \times  \frac{
  \sinh(u_j+u_k+\I \lambda) \sinh(u_j-u_k+\I \lambda)}
 {\sinh(u_j+u_k-\I \lambda) \sinh(u_j-u_k-\I \lambda)}.
\end{eqnarray}
with again $\epsilon = 0$ or $\frac{\pi}{2}$, depending on the
choice of boundary weights.

\section{Surface operator content}
\setcounter{equation}{0}

Let $\Lambda_0$ be the largest eigenvalue of the transfer matrix. 
From the explicit calculations on the loop models, we know that
the reduced free energy, $f_N = - N^{-1} \log \Lambda_0$, scales as
\beq
f_N = f_{\infty} + \frac{f_+}{N} + \frac{f_-}{N}
- \frac{\pi\zeta c}{24N^2} +o(N^{-2}) \; .  \label{c}
\eeq
Here $f_{\infty}$ is the bulk free energy, $f_\pm$ are surface free 
energies and 
$\zeta$ is a geometric factor. 
The rest of the eigenspectrum scales as
\beq
\log \frac{\Lambda_0}{\Lambda_\ell}=\frac{\pi\zeta (x_\ell + j)}{N} 
+ o(N^{-1})\, ,
\label{x}
\eeq
where $j=0,1,\ldots$ defines the conformal tower of eigenstates associated
with each eigenvalue.
Both (\ref{c}) and (\ref{x}) are in agreement with the expectation from 
conformal invariance \cite{ISZ,Cb,CH}. 
The quantities $f_{\infty}$ and $f_\pm$ are non-universal. 
Our interest here is in the universal scaling part of the spectrum
defined by the central charge $c$ and the surface scaling dimensions $x_\ell$.
Their values can be inferred from the underlying loop models.

\subsection{Dense $A_L$ model}

Given the above boundary weights, a comparison of the 
finite-size transfer matrix eigenspectra of the dense
$A_L$ model on the one hand, and the dense O($n$) loop model on the other,
reveals a selection rule on the parameter $m$ appearing in the
Bethe equations (\ref{denseBE}). This parameter labels the sectors 
of the transfer matrix of the vertex model representation of the
loop model. The restriction on $m$ arises from the mapping of height
configurations to arrow configurations on the underlying vertex model. 
Specifically, $m = \half N - \ell$,
for $\ell = 0,1, \ldots, \ell_{\rm max}$, where 
$\ell_{\rm max} = \lfloor \half (L-1) \rfloor$,  
the integer part of $\half (L-1)$.
We confirm numerically that the eigenspectrum of the face model
is exactly equivalent to the eigenspectrum of the loop model with
the above restriction. However, the equivalence is not one-to-one,
as the degeneracies differ. For example, the largest eigenvalue is
$L$-fold degenerate in the face model. There are
eigenstates in the
$\ell \le \ell_{\rm max}$ sectors of the loop model which
are not eigenstates of the face model. However,
these are precisely the same eigenvalues which appear in the sectors
with $\ell > \ell_{\rm max}$. As a result the 
surface operator content of the face model can be 
inferred from the known results for the loop model.  

The central charge and surface scaling dimensions of the loop
model are \cite{ob89,sb89,bosy95}
\beq
 c = 1 - \frac{6 \lambda^2}{\pi (\pi - \lambda)}, \qquad
x_\ell = \frac{\ell}{\pi}[\ell(\pi-\lambda) - \lambda] ,
\eeq
where $\ell$ is unrestricted.
The Bethe equation calculations also reveal a set of 
constant scaling dimensions, with $x = 2$. 
Now consider the face model.  
In terms of the Kac formula \cite{ISZ,Cb,CH}
\beq
\Delta_{r,s}={[h r-(h-1)s]^2-1\over 4h(h-1)}
\label{Delta}
\eeq
we have, recalling that $\lambda=\pi/h$ with $h=L+1$, 
\be
c&=&1-{6\over h(h-1)}, \label{cmin} \\
x_\ell&=&\Delta_{1,1+2\ell} , \label{xmin}
\quad \ell = 1, \ldots, \lfloor \half (h-2) \rfloor .
\ee
Along with the constant dimension $x=2$, this defines the
surface operator content of the dense $A_L$ face model. 

The first example is the Ising model with $h=4$, 
$c=\half$, $x_\sigma = \half$ and $x_\epsilon=2$. 
The next few cases are $h=5$, with 
$c=\frac{7}{10}$, $x = \frac{3}{5}$, $x=2$, and $h=6$,
$c=\frac{4}{5}$, $x = \frac{2}{3}$, $x=2$, $x = 3$.
Recall that the boundary heights are fixed along each boundary, 
but free to take all allowed values. The operator content induced by
other boundary configurations
and different geometries have also been considered for this particular
family of models \cite{sb89,c89,chim,pap}. 

The modular invariant partition functions can be constructed
from the central charge and conformal dimensions through
the Virasoro characters \cite{ISZ,Cb,CH} 
\be
\chi_{r,s} &=& \frac{q^{- \frac{c}{24} + \Delta_{r,s}}}{Q(q)} 
\sum_{n=-\infty}^\infty q^{n^2 h(h-1) + n h r} \left[ 
q^{-n(h-1)s} - q^{n(h-1)s + r s} \right], \nonumber \\
&=& q^{- \frac{c}{24} + \Delta_{r,s}} \sum_{n=0}^\infty 
d_n(\Delta_{r,s}) q^n,  
\ee
where
\beq
Q(q) = \prod_{n=1}^{\infty} (1-q^n) .
\eeq
The first few integer co-efficients $d_n(\Delta_{r,s})$ are tabulated
in, e.g., \cite{CH}. 
From the face model eigenspectra we see that the first few modular
invariant partition functions for the 
fixed boundaries under consideration are
\be
h=4, \quad Z &=& 3\chi_{1,1} + 3\chi_{1,3} \,,  \nonumber\\
h=5, \quad Z &=& 4\chi_{1,1} + 6\chi_{1,3} \,,  \\
h=6, \quad Z &=& 5\chi_{1,1} + 9\chi_{1,3} + 5 \chi_{1,5}. \nonumber
\ee
We can add $h=3, Z=2\chi_{1,1}$ \cite{sb89} to begin this list.
These relations define the asymptotic degeneracies of the eigenspectrum.

\subsection{Dilute $A_L$ model}

The transfer matrix eigenspectrum of the dilute $A_L$ face model is
equivalent to that of the dilute loop model with the restriction 
$m = N - \ell$, where $\ell = 0,1, \ldots, \ell_{\rm max}$, with 
$\ell_{\rm max} = L-1$. Again it is precisely the eigenvalues 
in the $\ell > \ell_{\rm max}$ sectors 
which do not appear in the eigenspectrum of the face model.

The central charge and surface scaling dimensions of the
dilute O($n$) model with boundary weights (\ref{dbw}) have been calculated
from the exact solution of section 2.2 at the particular value
$u=\lambda$ \cite{BS,by95a,by95b,yb95b}. This is the so-called honeycomb 
limit of relevance
to the surface critical behaviour of self-avoiding random 
walks \cite{DS,by95a,by95b,BO}. 
It is straightforward to extend
these calculations to the wider range $0 < u < 3\lambda$. Indeed,
the central charge and scaling dimensions are independent of the
spectral parameter $u$. The central charge is
\beq
c=1 - 6(g-1)^2/g,   \label{cdilute}
\eeq
where $\pi g = 2 \pi - 4\lambda$. Three sets of
``geometric'' scaling dimensions have been derived, each 
corresponding to a choice of the boundary weights. For $\epsilon=0$, 
the result is \cite{DS,BS,yb95b}
\beq
X_\ell^{\rm O - \rm O} = {\mbox{\small $\frac{1}{4}$}}g \ell^2 +
\half (g-1) \ell. 
\eeq
In the language of surface critical phenomena, O$-$O labels the  
ordinary surface transition. The choice $\epsilon=\frac{\pi}{2}$
(S$-$S) corresponds to the special surface transition. 
The other possibility
is mixed O$-$S boundary conditions with $\epsilon=0$ on
one side of the strip and $\epsilon=\frac{\pi}{2}$ on the other,
for which there is also a Bethe solution \cite{by95b,yb95b}. The 
extraordinary
surface transition has also been recently discussed \cite{BC}.
We confine our attention here to the O$-$O boundaries. The thermal
dimension has also been calculated, with in this case $x=2$.

For regime 1, $g = \frac{h}{h-1}$ with $h=L+2$. 
The central charge (\ref{cdilute}) 
gives the unitary minimal result (\ref{cmin}).
The scaling dimensions are given in terms of the Kac formula by
\beq
X_\ell^{\rm O - \rm O} = \Delta_{1+\ell,1} \, ,  \quad \ell=1,\ldots,h-3.
\eeq
As for the dense model, the
Ising case has $h=4$ with $c=\half$, 
$x_\sigma = \half$ and $x_\epsilon=2$.

The first few conformal partition functions are given by 
\be
h=4, \quad Z &=& 2\chi_{1,1}+2\chi_{2,1} \,,  \nonumber\\
h=5, \quad Z &=& 3\chi_{1,1}+4\chi_{2,1}+3\chi_{3,1}\,,  \\
h=6, \quad Z &=& 4\chi_{1,1}+6\chi_{2,1}+6\chi_{3,1}+4\chi_{4,1}\, . 
\nonumber
\ee
Since $\chi_{2,1} = \chi_{1,3}$ for $h=4$ both the dense and
dilute results are essentially the same for the Ising case.

For regime 2, $g = \frac{h-1}{h}$ with $h=L+1$. The central charge
is again given by (\ref{cmin}) with scaling dimensions 
\beq
X_\ell^{\rm O - \rm O} = \Delta_{1,1+\ell} \, , \quad \ell=1,\ldots,h-2.
\eeq
In this regime the Ising results for $h=4$ are 
$c=\half$, $x = \ffrac{1}{16}$, $x = \half$ and $x=2$. 
Here the first few conformal partition functions are given by 
\be
h=4, \quad Z &=& 3\chi_{1,1}+4\chi_{1,2}+3\chi_{1,3}\,,  \nonumber\\
h=5, \quad Z &=& 4\chi_{1,1}+6\chi_{1,2}+6\chi_{1,3}+4\chi_{1,4} \, , \\
h=6, \quad Z &=& 5\chi_{1,1}+8\chi_{1,2}+9\chi_{1,3}+8\chi_{1,4} +
                 5\chi_{1,5} \, ,
\nonumber
\ee
again from which the general pattern is readily apparent.

\section{Conclusion}

A set of particular fixed boundary weights has been constructed 
for both the dense and dilute $A_L$ face models from the
boundary weights of the related loop model. The transfer matrix
eigenspectra of the face models is seen to agree with that of
the loop models under restriction of the parameter $\ell$. 
The surface operator
content and the conformal partition functions then follow from the
results obtained via the Bethe equations for the loop models. 
It will be particularly worthwhile to generalise these Bethe
solutions in terms of the elliptic $\vartheta$-functions of the
off-critical $A_L$ models, as was done originally
in the bulk \cite{ABF,BNW}. 

\ack
It is a great pleasure to wish Professor J.M.J. van Leeuwen 
all the best on this occasion. I look back fondly on my first postdoc, 
in Leiden, under his inspirational guidance. Among other things,
he taught me to take my first steps on the ice.
This manuscript has benefited from helpful discussions with 
Katherine Seaton, Ole Warnaar and  Yu-kui Zhou.


\end{document}